\documentclass[aps,prl,reprint,longbibliography,superscriptaddress]{revtex4-1}

\usepackage{graphicx,amsmath,amssymb,color,amsfonts, bm}
\usepackage[dvipsnames]{xcolor}

\usepackage[colorlinks=true]{hyperref}  
\hypersetup{
    bookmarks=true,         
    unicode=false,          
    pdftoolbar=true,        
    pdfmenubar=true,        
    pdffitwindow=false,     
    pdfstartview={FitH},    
    pdftitle={},    
    pdfauthor={},     
    pdfsubject={},   
    pdfcreator={},   
    pdfproducer={}, 
    pdfkeywords={} {} {}, 
    pdfnewwindow=true,      
    colorlinks=true,       
    linkcolor=blue, 
    citecolor=blue,        
    filecolor=magenta,      
    urlcolor=blue,           
        breaklinks=true
} 

\newcommand{\be}{\begin{equation}}
\newcommand{\ee}{\end{equation}}
\newcommand{\bea}{\begin{eqnarray}}
\newcommand{\eea}{\end{eqnarray}}
\newcommand{\sect}[1]{\vspace{1mm}\noindent{\bf {#1}}.---}

\begin{document}

\title{Far-from-equilibrium universality in the two-dimensional Heisenberg model}

\author{Joaquin F. Rodriguez-Nieva}
\email{jrodrigueznieva@stanford.edu}
\affiliation{Department of Physics, Stanford University, Stanford, CA 94305, USA}

\author{Asier Pi\~neiro Orioli}
\affiliation{JILA, Department of Physics, University of Colorado, Boulder, CO 80309, USA}
\affiliation{Center for Theory of Quantum Matter, University of Colorado, Boulder, CO 80309, USA}

\author{Jamir Marino}
\affiliation{Institut f\"ur Physik, Johannes Gutenberg Universit\"at Mainz, D-55099 Mainz, Germany}

\begin{abstract}
We characterize the universal far-from-equilibrium dynamics of the two-dimensional quantum Heisenberg magnet isolated from its environment. For a broad range of initial conditions, we find a long-lived universal prethermal regime characterized by self-similar behavior of spin-spin correlations. We analytically derive the spatial-temporal scaling exponents and find excellent agreement with numerics using phase space methods. The scaling exponents are insensitive to the choice of initial conditions, which include coherent and incoherent spin states with values of total magnetization and energy in a wide range. Compared to previously-studied self-similar dynamics in non-equilibrium $O(n)$ field theories and Bose gases, we find qualitatively distinct scaling behavior originating from the presence of spin modes which remain gapless at long times and which are protected by the global SU(2) symmetry. Our predictions, which suggest a new non-equilibrium universality class, are readily testable in ultra-cold atoms simulators of Heisenberg magnets. 
\end{abstract}


\maketitle

Extending the paradigm of universality to far-from-equilibrium regimes is a current frontier in physics. A distinguishing feature of universal dynamics in complex systems is the emergence of self-similar behavior. Close to equilibrium, diverse and seemingly distinct models can be classified using symmetry and dimensionality into universality classes sharing the same self-similar scaling~\cite{1977HH}. Far from equilibrium, however, the system can break a symmetry concomitant with detailed balance~\cite{sieberer2015thermodynamic,aron2018non} and can therefore exhibit self-similar dynamics beyond conventional equilibrium paradigms. Celebrated examples include  turbulence~\cite{frisch_1995,zakharov1992}, ageing~\cite{calabrese2005ageing}, phase-ordering kinetics~\cite{1994Bray,biroli2010kibble,jelic2011quench}, KPZ scaling~\cite{1986kpz}, reaction-diffusion models, and percolation~\cite{tauber2014critical}.

Universality and self-similar relaxation typically arise in open systems with external friction and noisy forces that drive the system to or across a critical point~\cite{1977HH,tauber2014critical,sieberer2013dynamical,dalla2012dynamics,marino2016quantum,mitra2006nonequilibrium}. Remarkably, recent theoretical works have shown that self-similar scaling can also occur in isolated systems  where the system acts as its own  bath. Prominent examples of dynamical scaling in isolated quantum many-body systems include pre-thermal critical states~\cite{chandran2013equilibration,gagel2015universal,maraga2015aging,calabrese2005ageing,gambassi2011quantum,sciolla2013quantum,smacchia2015exploring,gagel2014universal,chiocchetta2016short,chiocchetta2017dynamical,chiocchetta2016universal,lerose2019impact,lerose2018chaotic} and non-thermal fixed points in scalar and gauge theories~\cite{Berges:2008wm,Micha:2004bv,NowakPRB84,Berges:2012us,NowakPRA85,PineiroPRD92,Karl:2016wko,GasenzerMikheevPRA99,SchmiedBlakie_PRA2019,WalzPRD97,ChantesanaPRA99,MoorePRD93,Berges:2016nru,Berges:2012iw,Berges:2013eia,Berges:2014bba,Boguslavski:2019fsb}. Research into these phenomena is further fueled by a surge of experimental evidence for universal dynamics in cold atoms~\cite{Prufer:2018hto,Erne:2018gmz,HadzibabicNature563,Prufer:2019kak,Zache:2019xkx,GliddenHadzibabic_Arxiv2020,smale2019observation}, and dynamical phase transitions in trapped ions~\cite{zhang2017observation} and cavity QED~\cite{muniz2020exploring}. 

Despite the ubiquity of spin models in condensed matter and AMO experiments, a systematic classification of non-equilibrium universality in such systems is still lacking. Indeed, non-equilibrium universality has mostly been explored analytically and numerically in models with $U(n)$ and $O(n)$ symmetries where self-similarity arises in the regime of large bosonic occupations~\cite{PineiroPRD92,Karl:2016wko,GasenzerMikheevPRA99,MoorePRD93,APO_PRL122,Boguslavski_PRDRap2020} or for quantum quenches in the $n\to \infty$ limit~\cite{gambassi2011quantum,chiocchetta2017dynamical,maraga2015aging,gagel2014universal,chandran2013equilibration}. Recent experiments in cold atomic gases have started to probe these non-equilibrium bosonic regimes~\cite{Prufer:2018hto,Erne:2018gmz,HadzibabicNature563,Prufer:2019kak,Zache:2019xkx,GliddenHadzibabic_Arxiv2020} and have also paved the way to explore other dynamical regimes in fully-tunable spin systems~\cite{2014spiralexp, 2020spiralketterle,jepsen2021transverse}, including tunable symmetries and the spatial dimension. Relevant to this new generation of experiments, here we show that the two-dimensional isotropic Heisenberg model at finite energy exhibits a non-thermal fixed point which is qualitatively distinct from previous instances of scaling, and we characterize its universal properties analytically and numerically.

Central to our discussion is the role of dimensionality $d=2$ and the presence of global SU(2) symmetry. First, the absence of a finite temperature symmetry-breaking phase transition in $d=2$ precludes scaling due to other well-studied phenomena like coarsening or ageing. Having two dimensions also bestows magnetization fluctuations a quasi-long-range character, which is an essential feature used to analytically compute the scaling exponents. Second, the global SU(2) symmetry constraints the nature of excitations in the system and their corresponding interactions. For instance, a recent work by one of us~\cite{bhattacharyya2020universal} showed that, close to the fully-polarized ferromagnetic ground state, the SU(2) symmetry constrains the interaction between magnons (the low-energy bosonic-like quasiparticles) and gives rise to slow magnon relaxation with anomalous scaling~\cite{bhattacharyya2020universal}. Here we show that the combination of symmetry and dimensionality alone, without any further assumptions like proximity to the fully-polarized ground state or bosonic approximations, can lead to a universal prethermal regime which is dominated by gapless spin modes and which is qualitatively distinct from previously-studied U($n$) and O($n$) theories in several important ways, as we discuss below. 

\begin{figure}
\centering\includegraphics[scale = 0.93]{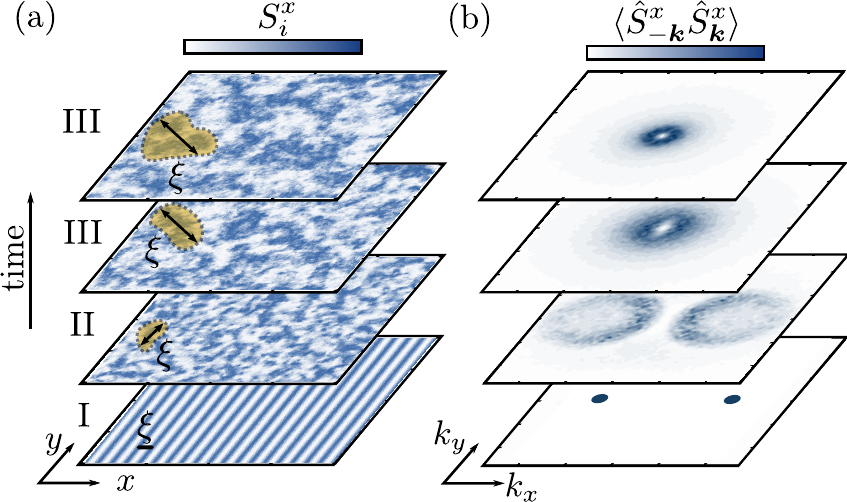}
\caption{Evolution of the spin-spin correlation function starting from an initially uncorrelated spin spiral state along the $\hat{x}$-direction for the two-dimensional Heisenberg model. The stages of relaxation are: (I) instability triggered by quantum fluctuations, (II) depletion of the macroscopically occupied state, (III) growth of magnetization fluctuations with a time-dependent correlation length $\xi$ (shown at two different times). Panel (a) shows a single semiclassical realization of spin $S_i^x$ configurations in real space, and panel (b) shows the (quantum) spin-spin correlation function in momentum space. }
\label{fig:schematics}
\end{figure}

More specifically, starting from an initial textured state with a characteristic wavevector $q$, we show that the equal-time spin-spin correlation functions exhibit self-similar scaling in a wide intermediate-time window,
\be
\sum_{a=x,y,z}\langle \hat{S}_{-\bm k}^a(t) \hat{S}_{\bm k}^a(t)\rangle = t^\alpha \Phi(t^{\beta}|{\bm k}|),
\label{eq:scaling}
\ee
for a broad range of initial conditions and arbitrary spin number $S$. The spatial-temporal scaling exponents $(\alpha,\beta)$ are independent of the details of the initial conditions, whereas the universal function $\Phi$ is only sensitive to a combination of $q$ and the global magnetization of the initial state. In a loose sense, the initial lengthscale of the spin texture $\xi = 1/q$ defines a dynamical renormalization group integration scale that governs the temporal scaling of correlation functions, see Fig.\,\ref{fig:schematics}. Using this physical picture, combined with analytical considerations based on symmetries and the structure of the equations of motion, we derive analytically the scaling exponents for a Gaussian and an interacting non-thermal fixed point and find excellent agreement with numerics using the truncated Wigner approximation (TWA)~\cite{spintwa1,spintwa2}. We also observe numerically the dynamical crossover from the Gaussian to the interacting non-thermal fixed point. 
Remarkably, our results, $\alpha=\beta d$ and $\beta=1/3$, agree within numerical uncertainties with the exponents found for magnon dynamics close to the FM state~\cite{bhattacharyya2020universal}, therefore suggesting the existence of a {\it single} non-thermal fixed point encompassing very broad energy and magnetization sectors. This contrasts with bosonic theories where different initial conditions can lead to different scaling regimes~\cite{NowakPRB84}. 
The present results, combined with those in a recent work by one of us~\cite{rodriguez2020turbulent} which found the asymptotic behavior $\Phi(x)\sim x^{-\nu_E}$ (with $\nu_E = 10/3$) using wave turbulence theory, allow us to fully characterize the universal spatial-temporal features of the non-thermal fixed point in terms of the three universal numbers $\alpha$, $\beta$, and $\nu_E$.

A key insight of the present work is that the intermediate-time self-similar dynamics is governed by gapless spin excitations whose gapless nature is protected by the global SU(2) symmetry. We demonstrate this numerically by computing unequal-time correlations~\cite{Boguslavski_PRDRap2020,APO_PRL122,BoguslavskiLappiPRD98,Schachner:2016frd,Schlichting_NPB2020}, where we find that the effective gap, or energy  
$\omega_{k}$ required to excite a long-wavelength mode on top of the prethermal state, remains zero at long times irrespective of the initial conditions or model parameters, as long as SU(2) is preserved. This circumstance is a hallmark of  our model. In contrast,  the effective gap to excite quasiparticles in O($n$) models out of equilibrium is not protected and typically grows due to the interplay between fluctuations and quartic interactions~\cite{PineiroPRD92,APO_PRL122,Boguslavski_PRDRap2020,chiocchetta2017dynamical,chandran2013equilibration,maraga2015aging,chiocchetta2017dynamical,smacchia2015exploring}. When SU(2) is reduced to U(1) in our model, we find different scaling exponents, therefore reinforcing the distinction between non-equilibrium universality in the isotropic Heisenberg model and previous instances of scaling in U(1) models.

In the spirit of the Halperin-Hohenberg classification~\cite{1977HH}, the values of $(\alpha,\beta)$ that we find, combined with the presence of an additional slow mode with quadratic dispersion, suggest that the Heisenberg model belong to a different non-equilibrium universality class than Bose gases~\cite{PineiroPRD92,Karl:2016wko,GasenzerMikheevPRA99,SchmiedBlakie_PRA2019,APO_PRL122,Boguslavski_PRDRap2020,MoorePRD93}, dissipative $O(n)$ models~\cite{1975belavinpolyakov,1997rutenberg,1995rutenbergbray,1995spintextures, 2000spintextures}, and other non-integrable high dimensional spin systems~\cite{2015PRX-Babadi,bhattacharyya2020universal,2020spinspiral,rodriguez2020turbulent,rodriguez2020hydrodynamic,PhysRevLett.122.127202,PhysRevB.98.220303,PhysRevB.101.180302,PhysRevLett.119.220604,PhysRevB.97.045407}, as we discuss below. This is quite surprising in light of the well-known similarities between the Heisenberg model and O(3) scalar models at equilibrium. We emphasize, however, that the scaling regime discussed in the present work is intrinsically different from the predictions of the Halperin-Hohenberg classification: while the latter describes universal behaviour close to thermodynamic equilibrium, here we consider a dynamical regime where equilibrium properties, such as the fluctuation-dissipation relation, are violated. 

\sect{Microscopic model}We consider the two-dimensional isotropic Heisenberg model on a square $L\times L$ lattice with lattice constant $\ell$ and total number of sites $N=L^2$: 
\be
\hat{H} = -J\sum_{\langle i,j \rangle} \left( \hat{S}_i^x\hat{S}_j^x+\hat{S}_i^y\hat{S}_j^y+  \hat{S}_{i}^z\hat{S}_j^z \right),
\label{eq:hamiltonian}
\ee
where $\langle i , j \rangle$ denotes summation over nearest neighbours. Each site has a spin $S$ degree of freedom and periodic boundary conditions are used in each spatial direction. This model has an SU(2) symmetry with conserved global spin magnetization $S^a\equiv \sum_i \langle \hat{S}^a_i \rangle$ for $a\in\{x,y,z\}$, where $\langle \cdot \rangle$ denotes the quantum expectation value. 

As initial condition, we consider an uncorrelated pure product state where the spins form a spiral in real space parametrized by a wavevector $\bm q$ and angle $\theta$, 
\be
\langle \hat{S}_i^\pm(0)\rangle = S\sin\theta e^{\pm i{\bm q}\cdot{\bm r}_i},\quad \langle \hat{S}_i^z(0) \rangle = S\cos\theta, 
\label{eq:initial}
\ee
with $\hat{S}_i^\pm = \hat{S}_i^x\pm i \hat{S}_i^y$ and ${\bm r}_i$ the position of site $i$. Other classes of initial conditions are considered in the Supplemental Information (SI). Equation (\ref{eq:initial}) defines a characteristic timescale $1/\tau_* = JS^2\sin^2\theta[2-\cos(q_x\ell)-\cos(q_y\ell)]$ associated to the energy density of the initial state. 

\sect{Spatial-temporal scaling via phase space methods}We begin by computing the real time dynamics using TWA~\cite{polkovnikov2010phase}. This method incorporates quantum fluctuations present in the initial state by considering classical spins ${\bm S}_i=(S^x_i,S^y_i,S^z_i)$ with quantum noise at $t=0$ and evolving them with the classical Landau-Lifshitz equations of motion. Defining ${\bm S}_i^\perp$ as the transverse magnetization to $\langle \hat{\bm S}_i(0) \rangle$ for the initial condition (\ref{eq:initial}), we assume initial Gaussian fluctuations of ${\bm S}_i^\perp$ given by $\langle {\bm S}_i^\perp\rangle_\text{cl} = 0$ and $\langle {\bm S}_i^\perp\cdot {\bm S}_i^\perp\rangle_\text{cl} = S$, where $\langle \cdot \rangle_\text{cl}$ denotes average over classical trajectories.

The subsequent dynamics follows three relaxation stages as illustrated in Fig.~\ref{fig:schematics}: (I) quantum fluctuations trigger a dynamical instability which leads to (II) a depletion of the macroscopically occupied state $\mathbf{q}$ and is followed by (III) a scaling regime in which magnetization fluctuations
 grow with a time-dependent correlation length $\xi(t)\sim t^{\beta}$ and which exhibits self-similar evolution following Eq.~(\ref{eq:scaling}).
 
\begin{figure}
\centering\includegraphics[scale = 0.93]{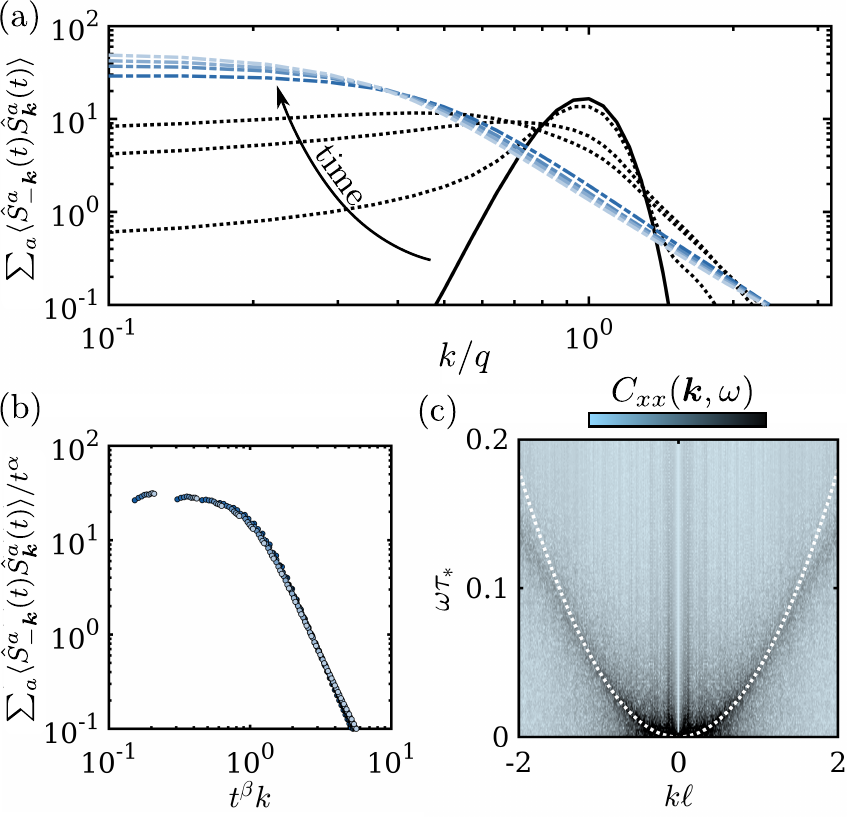}
\vspace{-2mm}
\caption{(a) Evolution of the spin-spin correlation function $\sum_a \langle \hat{S}_{-\bm k}^a\hat{S}_{\bm k}^a\rangle$ after a quench from an initial spin spiral. Shown with solid and dotted lines are the initial distribution and the process of depletion of the initial $q$ mode (stages I-II), respectively. We show with colored dashed lines the spin-spin correlation in the self-similar regime (stage III). Dotted lines are plotted for $t/\tau_* = 3,4,5$, and dashed lines are plotted for the range $15 < t/\tau_* < 40$, with lighter colors for increasing $t$. (b) Collapsed datapoints during stage III, c.f. Eq.~(\ref{eq:scaling}),   with scaling exponents $\alpha = \frac{2}{3}$ and $\beta  = \frac{1}{3}$. (c) Fourier transform of the unequal time spin-spin correlation function $C_{xx}({\bm k},\omega) \equiv \int dt\, e^{i\omega t} \langle \frac{1}{2} \{ \hat{S}^x_{-\bm k}(t_0+t), \hat{S}^x_{\bm k}(t_0) \} \rangle$ for $t_0 = 20\tau_*$. Different values of $t_0$ do not affect the qualitative features of the plot. Shown with doted lines is a quadratic fit of the mode dispersion. Simulations parameters: $L=800$, $q_x = 0.2$, $\theta = \pi/2$, $S=5$.}
\label{fig:fullspiral}
\end{figure}

Figure \ref{fig:fullspiral}(a) shows the evolution of the spin-spin correlation function $\sum_a \langle \hat{S}_{-\bm k}^a(t) \hat{S}_{\bm k}^a(t)\rangle$~\cite{twafootnote} during the three stages of relaxation for a full spiral ($\theta=\pi/2$).
Here, $\hat{S}^a_{\bm k}=\frac{1}{\sqrt{N}}\sum_i e^{i {\bm k}\cdot {\bm r}_i} \hat{S}^a_i$ denotes the discrete Fourier transform with wavevector ${\bm k}$.
The decay of the initial macroscopically occupied mode ${\bm q}$ (solid line) occurs on a timescale of approximately $4\tau_*$~\cite{2020spinspiral} and leads to a quick redistribution of fluctuations into other ${\bm k}$ modes [stages I and II in Fig.~\ref{fig:schematics}; dotted lines in Fig.~\ref{fig:fullspiral}(a)]. At later times [stage III in Fig.~\ref{fig:schematics}; blue dashed lines in Fig.~\ref{fig:fullspiral}(a)] the system exhibits self-similarity as demonstrated in Fig.~\ref{fig:fullspiral}(b) by the excellent collapse of the rescaled curves with 
\be
    \alpha = 0.63\pm 0.05, \quad \beta = 0.34\pm 0.03,
\label{eq:alphabetanumerics}
\ee
which also agree with our analytical estimates below (the procedure for fitting the exponents is discussed in the SI). 

The SU(2) symmetry of the Heisenberg model precludes the opening of an effective gap during the dynamics.
To find the relevant excitations, we evaluate in Fig.~\ref{fig:fullspiral}(c) the unequal time spin-spin correlation functions in frequency space, $C_{xx}({\bm k},\omega) \equiv \int dt\, e^{i\omega t} \langle \frac{1}{2} \{ \hat{S}^x_{-\bm k}(t_0+t), \hat{S}^x_{\bm k}(t_0) \} \rangle$~\cite{twafootnote}, for intermediate times $t_0$. This shows that the self-similar dynamics is governed by gapless excitations at all time scales $t_0$, even when the initial state is far from 
the fully-polarized ground state. The dispersion of this mode is consistent with $\omega \sim k^2$ (white dotted line) for wavevectors $k\ell\gtrsim 0.08$ below which finite-size effects become sizable. 

\sect{Derivation of the scaling exponents $(\alpha,\beta)$}We now analytically estimate the scaling exponents  assuming for simplicity that the system has no net magnetization (non-zero magnetization does not affect the argument in any essential way). In this case, spin-spin fluctuations eventually become isotropic both in real and spin space. We find this to occur after a short transient timescale $\approx 5\tau_*$ (c.f.~Fig.~\ref{fig:schematics} and SI) and, therefore, the three components of the spin-spin correlation function exhibit the same scaling. In addition, we find that the mean-field components $\langle \hat{S}^a_{\bm k}(t)\rangle$ vanish on average at the onset of stage III. As a result, we use the full and connected component of $\langle \hat{S}^a_{-\bm k}(t) \hat{S}^a_{\bm k}(t)\rangle$ indistinguishably. 

The first relation between $\alpha$ and $\beta$ is obtained from the local constraint of spin operators $\hat{\bm S}_i\cdot\hat{\bm S}_i = S(S+1)$. In momentum space, this relation is written as
\be
\frac{1}{N}\sum_{{\bm k},a} \hat{S}_{-\bm k}^{a}\hat{S}_{\bm k}^{a} = S(S+1).
\label{eq:S2}
\ee
Equation (\ref{eq:S2}) is an exact relation that is independent of the state of the system. If the spin-spin correlation function satisfies Eq.~(\ref{eq:scaling}), then Eq.~(\ref{eq:S2}) implies that $t^{\alpha-d\beta}\int \frac{d^d{\bm x}}{(2\pi)^d}\Phi(|{\bm x}|)$ is a constant or, equivalently, that $\alpha$ and $\beta$ are related through
\be
\alpha = d \beta.
\label{eq:relation1}
\ee
Interestingly, we note that the relation (\ref{eq:relation1}) also appears in other scaling regimes of different microscopic nature. For instance, Eq.(\ref{eq:relation1}) appears in coarsening dynamics when one assumes that correlations are governed by a single lengthscale given by the typical size of the ordered regions [c.f., Eq.(7) of Ref.~\cite{1994Bray}]. It also appears in the universal dynamics of a Bose gas~\cite{PineiroPRD92}, where Eq.~(\ref{eq:relation1}) is equivalent to boson number conservation in the self-similar range. Here, instead, Eq.(\ref{eq:relation1}) is a consequence of the constraint on spin length. 

\begin{figure*}[t]
\centering\includegraphics[scale = 1.0]{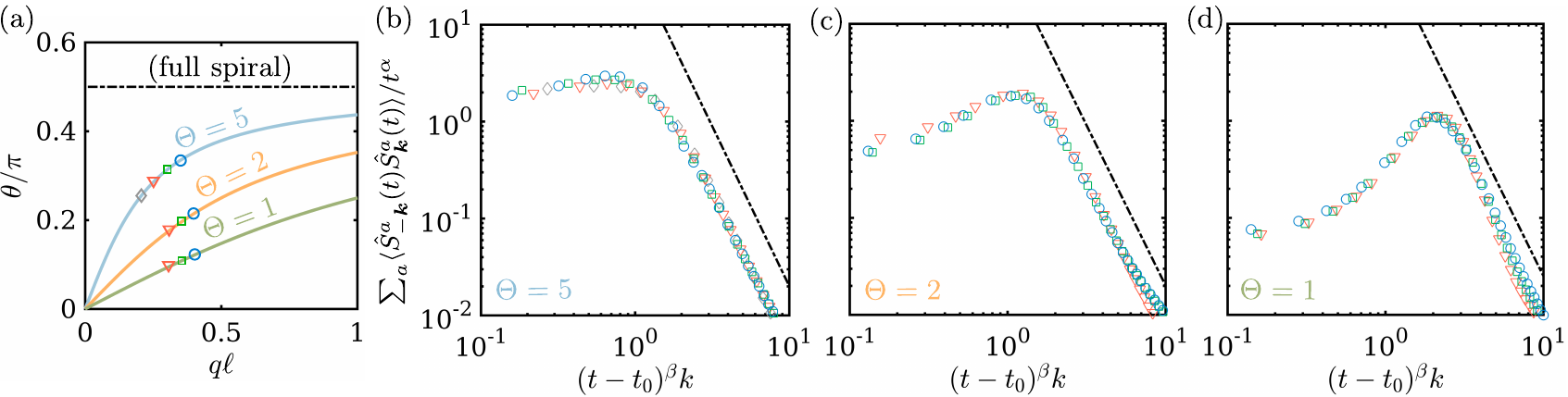}
\vspace{-2mm}
\caption{Self-similar scaling of the spin-spin correlation functions shown for  initial conditions with the same value of $\Theta = \tan\theta/(q\ell)$. In panels (b)-(d), different symbols denote initial conditions with the different values of $(q,\theta)$ shown in panel (a). When collapsing the datapoints, we allow for a finite time shift $t_0$ to account for the initial-condition-dependent dephasing time. Shown with dashed-dotted lines is the universal power law scaling $\sim k^{-10/3}$ associated to spin turbulence, see Ref.~\cite{rodriguez2020turbulent}.}
\label{fig:rscaling}
\end{figure*}

The second relation between $\alpha$ and $\beta$ is obtained from dynamics. The time dependence of the spin-spin correlation function $\langle \hat{S}_{-\bm k}^a \hat{S}_{\bm k}^a\rangle$ can be straightforwardly obtained from the microscopic equations of motion of the spin operators, $\partial_t \hat{S}_{i}^a = \sum_{j,b,c} \epsilon_{abc} \hat{S}_i^b \hat{S}_j^c$, which yields 
\be
\partial_t \langle \hat{S}_{-\bm k}^a \hat{S}_{\bm k}^a \rangle = 2 \sum_{{\bm p},b,c} (\gamma_0-\gamma_{\bm p}){\rm Re}\left[\epsilon_{abc}\langle \hat{S}_{-\bm k}^a\hat{S}_{\bm k - \bm p}^b\hat{S}_{\bm p}^c\rangle\right],
\label{eq:S2eom}
\ee
where $\gamma_{\bm p} = \sum_{\bm \ell}e^{i{\bm p}\cdot{\bm \ell}}$ (${\bm \ell}$ are unit cell vectors) (see details in the SI). The central assumption in our derivation is that a single lengthscale $\xi$ governs the scaling of two and three point correlation functions. The validity of such stringent assumption can be justified from the long-range character of spin modes and the absence of other characteristic lengthscales in $d=2$ (e.g., a non-zero effective gap or localized defects). This translates to modes being macroscopically and democratically occupied within a region $\bm k\lesssim 1/\xi$, contrary to a conventional condensate with long-range order where a single mode $\bm k=0$ is macroscopically occupied. Such quasi-long-range character in low-dimensional system is often referred to as a quasi-condensate in the BEC literature~\cite{Castin_PRA2003}. In addition, the rate of change of $\xi$ is assumed to be much slower than the microscopic timescale $\tau_*$ defined above, such that fast microscopic fluctuations can be integrated out. As a result, dimensional analysis of Eq.~(\ref{eq:scaling}) suggests that ${\rm Re}\left[\epsilon_{abc}\langle \hat{S}_{-\bm k}^a\hat{S}_{\bm k - \bm p}^b\hat{S}_{\bm p}^c\rangle\right] \sim \xi^{3\alpha/2\beta}\Psi({\bm k}\xi,\bm p\xi)$, with $\xi\sim t^\beta$ and $\Psi$ a scaling function. Using this scaling form in the right-hand side of Eq.~(\ref{eq:S2eom}) in the long wavevelength limit [i.e., taking $\gamma_0-\gamma_{\bm q}\approx ({\bm q}\ell)^2$] yields $\xi^{3\alpha/2\beta-d-2}\Psi'({\bm k}\xi)$, with $\Psi'({\bm x}) = \int\frac{d^d{\bm y}}{(2\pi)^d}{\bm y}^2\Psi({\bm x},{\bm y})$. In addition, the left-hand-side of Eq.~(\ref{eq:S2eom}) scales as $\xi^{\alpha/\beta-1/\beta}$. Equating the scaling form on both sides of \eqref{eq:S2eom} results in 
\be
2(d+2)\beta = \alpha +2. 
\label{eq:relation2}
\ee
Combining~\eqref{eq:relation2} and~\eqref{eq:relation1} in $d = 2$ yields $\alpha = \frac{2}{3}$ and $\beta = \frac{1}{3}$, consistent with the numerical results in Eq.~(\ref{eq:alphabetanumerics}) and with the scaling found numerically within kinetic theory close to the ferromagnetic ground state  with $\theta\approx0$~\cite{bhattacharyya2020universal} (we note that kinetic theory failed to analytically yield the correct exponents from scaling arguments alone). We emphasize again that Eq.~(\ref{eq:relation2}) is only valid in $d=2$ as it relies on spin modes having a long-range character (in $d=3$, for example, the initial spin texture will collapse into localized skyrmions giving rise to qualitatively different prethermal dynamics). 

\sect{The universal scaling function}So far, we considered the dynamics of a single dimensionless parameter ${\xi}/{\ell}$ that governs the self-similar scaling. In addition, we empirically find that the scaling function $\Phi$ in Eq.(\ref{eq:scaling}) is sensitive to the magnetization sector of the state through the dimensionless ratio
\be
\Theta = \frac{\tan \theta}{q\ell},
\label{eq:Theta}
\ee
as shown in Fig.~\ref{fig:rscaling}. If $\theta = \pi/2$, then $\Theta\rightarrow\infty$ and the scaling function becomes independent of $q$. For other values of $\theta$ and $q$, we find an excellent collapse of the data points.

We note that the dimensionless parameter $\Theta$ essentially quantifies the ratio between magnetization fluctuations and the global magnetization, as we discuss next. The average magnetization is given by $S^z/N = S\cos\theta$. The average amplitude of magnetization fluctuations at the onset of stage III is obtained from the identity (\ref{eq:S2}) after removing the disconnected component associated to the global magnetization, which scales as $S^2\cos^2\theta$, and assuming that fluctuations are equally distributed in a phase space region of size $|{\bm k}| \lesssim q$. For large $S$, this results in ${\sum_a\langle S_{-\bm k}^aS_{\bm k}^a\rangle_{\rm c}}\sim (S\sin\theta q\ell)^2$, where `c' stands for connected. The ratio between these two quantities gives the empirical Eq.~(\ref{eq:Theta}). 

\sect{Dynamical crossovers}The self-similar scaling regime described above cannot be captured by an effective Gaussian description. This can be readily checked using a self-consistent Holstein-Primakoff approximation in the low spin-wave density expansion~\cite{halperin1969hydrodynamic,rodriguez2020hydrodynamic}, which would instead yield $\alpha=1$ and $\beta=1/2$ for the $d=2$ Heisenberg model (see SI). Interestingly, these Gaussian exponents can be observed at early times before crossing over into the non-thermal fixed point. An instance of this dynamical crossover, which is reminiscent of the phenomenon of prescaling~\cite{MazeliauskasBerges_PRL2019,SchmiedGasenzer_PRL2019}, is reported in the  SI. Therefore, two dynamical crossovers occur in the Heisenberg model: the first accompanies dynamics from a Gaussian to a non-Gaussian fixed point, while the second dictates the approach to the thermal state, which is the fate of dynamics for generic high dimensional non-integrable systems. 

\sect{Discussion}The self-similar exponents and scaling functions obtained here are distinct from those found in previous works on non-equilibrium dynamics in classical and quantum $O(n)$ field theories and $U(n)$ bosonic models~\cite{PineiroPRD92,Karl:2016wko,GasenzerMikheevPRA99,SchmiedBlakie_PRA2019,APO_PRL122,Boguslavski_PRDRap2020}. Compared to these previous works, the central difference of our results is that the dynamics of the Heisenberg model is dominated by gapless spin excitations at all times. This is in sharp contrast to typical non-thermal fixed points in bosonic theories, where an effective gap 
has been observed to be dynamically generated by fluctuations~\cite{PineiroPRD92,APO_PRL122,Boguslavski_PRDRap2020}.
This effective gap has been shown to lead to a modified (non-relativistic) effective theory at low momenta, which is characterized by different scaling exponents than the gapless theory. 
Our results also differ from scaling dynamics of $O(n)$ theories quenched to (or across) a critical point~\cite{chandran2013equilibration,maraga2015aging,chiocchetta2017dynamical,smacchia2015exploring}, where a dynamically generated gap only vanishes asymptotically in time: in these works, self-similarity occurs only if parameters and initial conditions are fine-tuned so as to guarantee a vanishing late-time effective gap. Compared to these bosonic theories, the interaction vertex of the Heisenberg model is `soft' after a spin wave expansion~\cite{rodriguez2020hydrodynamic,halperin1969hydrodynamic}, i.e.~it contains spatial gradients that are absent in $O(n)$ models, see SI. This  already sets a difference at the level of canonical power counting, and   suggests  that the two models cannot belong to the same universality class (see SI).

To further demonstrate the importance of the SU(2) symmetry, we show in the SI that, by adding anisotropic exchange $\hat{H} = \delta J_z\sum_{\langle i,j \rangle}\hat{S}_i^z\hat{S}_j^z$ to Eq.~(\ref{eq:hamiltonian}), the dynamical exponents flow to a different non-thermal fixed point. In particular, for the easy-plane case ($\delta J_z>0$) wherein the global symmetry is instead $U(1)$, we obtain the same scaling exponents $\alpha=1$, $\beta=1/2$ as those observed in bosonic $O(n)$ and $U(n)$ theories~\cite{PineiroPRD92,Karl:2016wko,MoorePRD93,GasenzerMikheevPRA99}. 

\sect{Conclusions}Our results on self-similar relaxation in    
the Heisenberg model extend to spin systems the paradigm  of scaling in the proximity of non-thermal fixed points, which were   so far thoroughly studied only for interacting bosonic and gauge theories~\cite{Berges:2008wm,Micha:2004bv,NowakPRB84,Berges:2012us,NowakPRA85,PineiroPRD92,Karl:2016wko,GasenzerMikheevPRA99,SchmiedBlakie_PRA2019,WalzPRD97,ChantesanaPRA99,MoorePRD93,Berges:2016nru,Berges:2012iw,Berges:2013eia,Berges:2014bba,Boguslavski:2019fsb}. Cold atoms  experiments have so far explored far-from-equilibrium transport and relaxation of one-dimensional quantum Heisenberg models~\cite{2020spiralketterle,2014spiralexp}. We believe that our results provide strong motivation to extend quantum simulators of spin models to higher dimensionality, where integrability is less prominent in constraining quantum dynamics. Another natural next step towards implementations would consist in considering long-range spin interactions $\propto 1/r^\zeta$, with the perspective to investigate the dependence of dynamical scaling exponents     with $\zeta$ (see Ref.~\cite{dauxois2002dynamics} for an equilibrium counterpart).

Our work opens up at least two interesting research avenues to explore. On the one hand, we have found scaling exponents which are remarkably robust to initial conditions belonging to different energy and magnetization sectors. It would be interesting to investigate whether this applies to other spin models with, e.g., SU($n$) symmetric interactions, which can also be studied in cold atom experiments~\cite{Gorshkov2010NatPh}.
On the other hand, our work has highlighted the essential role played by the symmetry-protected gaplessness of spin excitations. This poses the question of whether the scaling exponents observed in this work can also be reproduced in O($n$) models under conditions that might have been overlooked so far, or whether the Heisenberg spin model far from equilibrium is fundamentally different from O($n$) models. The latter would clash with equilibrium common wisdom and reinforce the intuition that non-equilibrium scaling is governed by intrinsically different mechanisms than equilibrium universality. 

\vspace{2mm}

{We are grateful to J.~Berges, K.~Boguslavski, E. Demler, T.~Gasenzer, and P. Glorioso for helpful discussions and collaboration on related topics. JFRN acknowledges the Gordon and Betty Moore Foundation’s EPiQS Initiative through Grant GBMF4302 and GBMF8686, the 2019 KITP program {\it Spin and Heat Transport in Quantum and Topological Materials}, and the National Science Foundation under Grant No. NSF PHY-1748958. JM acknowledges support by the Dynamics and Topology Centre funded by the State of Rhineland Palatinate.}


%

\clearpage

\renewcommand{\thefigure}{S\arabic{figure}}
\renewcommand{\theequation}{S\arabic{equation}}
\renewcommand{\thesection}{S\arabic{section}}
\setcounter{page}{1}
\setcounter{equation}{0}
\setcounter{figure}{0}
\setcounter{section}{0}

\begin{widetext}

\begin{center}
{\large\bf {SUPPLEMENTAL MATERIAL} \\ Far-from-equilibrium universality in the two-dimensional Heisenberg model}

\vspace{4mm}

Joaquin F. Rodriguez-Nieva$^1$, Asier Pi\~neiro Orioli$^{2,3}$, Jamir Marino
$^4$

{\small\it $^1$Department of Physics, Stanford University, Stanford, CA 94305, USA}

{\small\it $^2$JILA, Department of Physics, University of Colorado, Boulder, CO 80309, USA}

{\small\it $^3$Center for Theory of Quantum Matter, University of Colorado, Boulder, CO 80309, USA}

{\small\it $^4$Institut f\"ur Physik, Johannes Gutenberg Universit\"at Mainz, D-55099 Mainz, Germany}

\end{center}


\vspace{5mm}

The outline of the Supplemental Material is as follows. In Sec.\,I, we numerically demonstrate the insensitivity of the scaling exponents $(\alpha,\beta)$ to variations in the initial conditions by considering incoherent initial conditions, which are different from the spin spiral initial conditions used in the main text. In Sec.\,II, we study the time evolution of the spin-spin correlation functions for the transverse and longitudinal components of magnetization and show that they become isotropic even when the initial condition is not. In Sec.\,III, we show the fitting scheme used to numerically estimate the scaling exponents $(\alpha,\beta)$. In Sec.\,IV, we discuss the scaling exponents in the XXZ model. 
In Sec.\,V, we discuss the derivation of Eq.~(7) in the main text. In Sec.\,VI, we derive the canonical power counting used to estimate the scaling exponents of the gaussian fixed point and, in Sec.VII, we discuss the differences in the vertex interactions between the Heisenberg model and $O(n)$ models. 

\section{I. Independence of initial conditions}

To illustrate the universal character of our results, we repeat our calculations for incoherent initial conditions and show that the system exhibits the same scaling behavior as the one observed in the main text. Here we use initial conditions of the form 
\be
\langle S_i^x \rangle = S\cos\phi_i\sin\theta, \quad\langle S_i^y \rangle = S\sin\phi_i\sin\theta, \quad \langle S_i^z \rangle = S\sin\theta, \quad \phi_i = {\rm arg}\left[\sum_{\bm k} f_{\bm k}e^{i{\bm k}\cdot {\bm r}_i}\right],
\label{eq:incoherent}
\ee
where $f_{\bm k}$ is a Gaussian-distributed complex function $f_{\bm k} = e^{i\theta_{\bm k}-(|{\bm k}|-q)^2/\sigma_q^2}$ that satisfies $f_{-\bm k}=f_{\bm k}^*$, with $\sigma_q = 0.1 q$ and $\theta_{\bm k}$ a random phase uniformly sampled $[0,2\pi)$. Figure\,\ref{fig:incoherent}(a) shows the self-similar scaling for initial conditions with zero net magnetization and with finite magnetization computed through the Truncated Wigner Approximation (TWA). In both cases, we rescaled the datapoints using $(\alpha,\beta) = (2/3,1/3)$. We find excellent agreement with the results reported in the main text using spin spiral initial conditions. In addition, the scaling function also matches remarkably well with the ones found in the main text, both with zero magnetization (Fig.\,2 of main text) and non-zero magnetization (Fig.\,3 of main text).

\section{II. Isotropic spin-spin correlations}

In the analytical derivation of the scaling exponents in the main text, we assumed that the spin-spin correlation function becomes isotropic in spin space. Although this is a plausible assumption, we numerically check that this is indeed the case. Figures \ref{fig:incoherent}(b-c) show the evolution of the (b) $xx$ and (c) $zz$ spin-spin correlation function at short times computed using TWA. We find that in a short timescale $\approx 5\tau_*$ the spin-spin correlations become isotropic. In addition, both the $xx$ and $zz$ correlations exhibit the same scaling behavior. We recall that $\tau_*$ is defined from the initial conditions as $1/\tau_* = JS^2\sin^2\theta[2-\cos(q_x\ell)-\cos(q_y\ell)]$, with $(q_x,q_y)$ the characteristic wavevector of the initial state, and $\theta$ defines the magnetization of the initial states, $S^z = NS\cos\theta$. 

\begin{figure}
\centering\includegraphics[scale = 1.0]{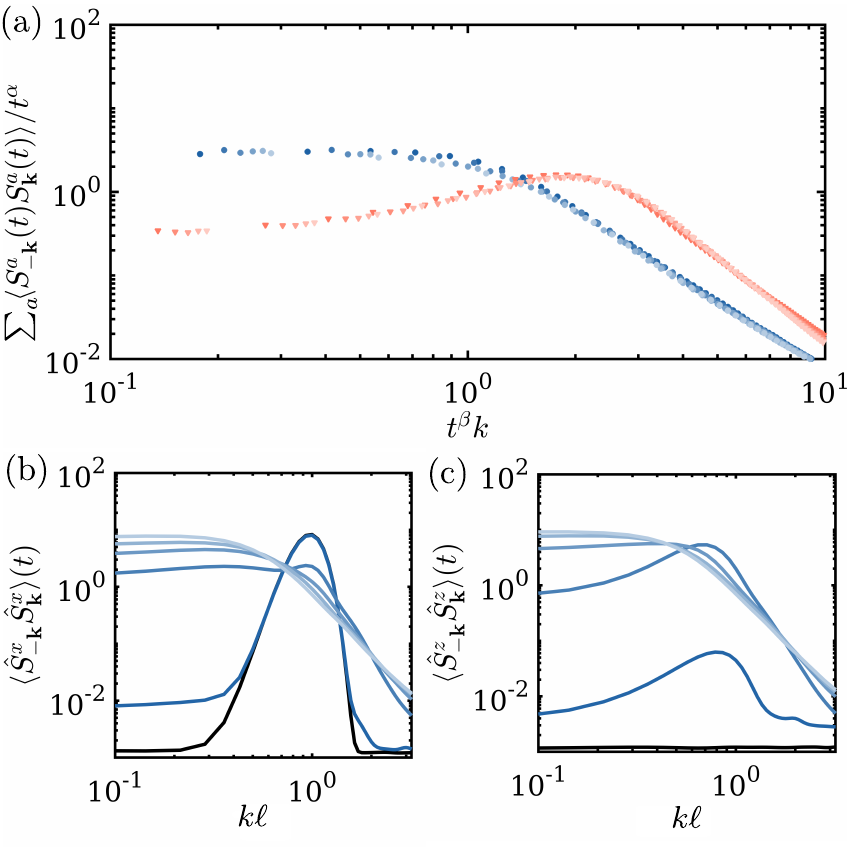}
\vspace{-2mm}
\caption{(a) Insensitivity of the scaling function and scaling exponents to the initial conditions. We show the rescaled spin-spin correlator $\sum_a \langle \hat{S}_{-\bm k}^a(t)\hat{S}_{\bm k}^a(t)\rangle$ using incoherent initial conditions [see Eq.(\ref{eq:incoherent})] with no net magnetization (blue circles) and with average magnetization $\langle S_i^z\rangle = S/2$ (orange triangles). Shown are datapoints in the time range $15\tau_*<t<40\tau_{*}$, with decreasing shades of color indicating increasing time. We rescaled the data with scaling exponents $(\alpha,\beta)=(2/3,1/3)$. (b-c) Evolution of the spin-spin correlation function $\langle \hat{S}_{-\bm k}^a\hat{S}_{\bm k}^a\rangle$ for (b) $a=x$ and (c) $a = z$. Shown with black lines is $\langle \hat{S}_{-\bm k}^a(0)\hat{S}_{\bm k}^a(0)\rangle$, and with blue lines is $\langle \hat{S}_{-\bm k}^a(t)\hat{S}_{\bm k}^a(t)\rangle$ for $t = 2\tau_*$ to $t=10\tau_*$ in steps of $2\tau_*$ (color shade decreases with time). The plots show that, after a transient time $\sim 5 \tau_*$, correlations become isotropic in spin space.}
\label{fig:incoherent}
\end{figure}

\section{III. Fitting of the scaling exponents}

To find the scaling exponents $(\alpha,\beta)$ that best fit the numerical data, we minimize an error function that quantifies the accuracy of the datapoint collapse, as we describe next. 
We denote as $C(|{\bm k}|,t) = \langle \hat{S}_{-\bm k}(t)\hat{S}_{\bm k}(t)\rangle_{\rm c}$ the connected component of the spin-spin correlation function. The values of $|\bm k| = k_i$ take discrete values on a lattice, and we choose different times $t_m$ within a time window $[t_0-\Delta t/2, t_0 + \Delta t/2]$ which shows self-similar scaling. Fixing the value of $(\alpha,\beta)$, we define the auxiliary variables 
\be
y_{i,m} = t_m^\alpha C(k_i,t_m), \quad x_{i,m} = t_m^\beta k_i.
\ee
For each value of $t=t_m$, these functions define an implicit function $y_m(x)$ which we can approximate by linearly interpolating the datapoints $(x_{i,m}, y_{i,m})$. As such, the error function can be defined as 
\be
{E}(\alpha,\beta) = \sum_{m,m'} \int dx|y_{m}(x)-y_{m'}(x)|.
\label{eq:error}
\ee

The value of $E(\alpha,\beta)$ for different values of $t_0$ and $\Delta t = 10\tau_*$ is plotted in Fig.~\ref{fig:errors}. For small $t_0 = 15\tau_*$, we find that the self-similar scaling can be fitted with $(\alpha,\beta) = (1,0.5)$. This is consistent with a Gaussian fixed point, see Sec.~V below. For $t_0 > 20\tau_*$, we find that the self-similar scaling can be best fitted with $(\alpha,\beta) = (2/3,1/3)$. The temporal dependence of the optimal fit is shown in Fig.~\ref{fig:betat} and shows the crossover from Gaussian to non-thermal fixed point. 

\begin{figure}
\centering\includegraphics[scale = 1.0]{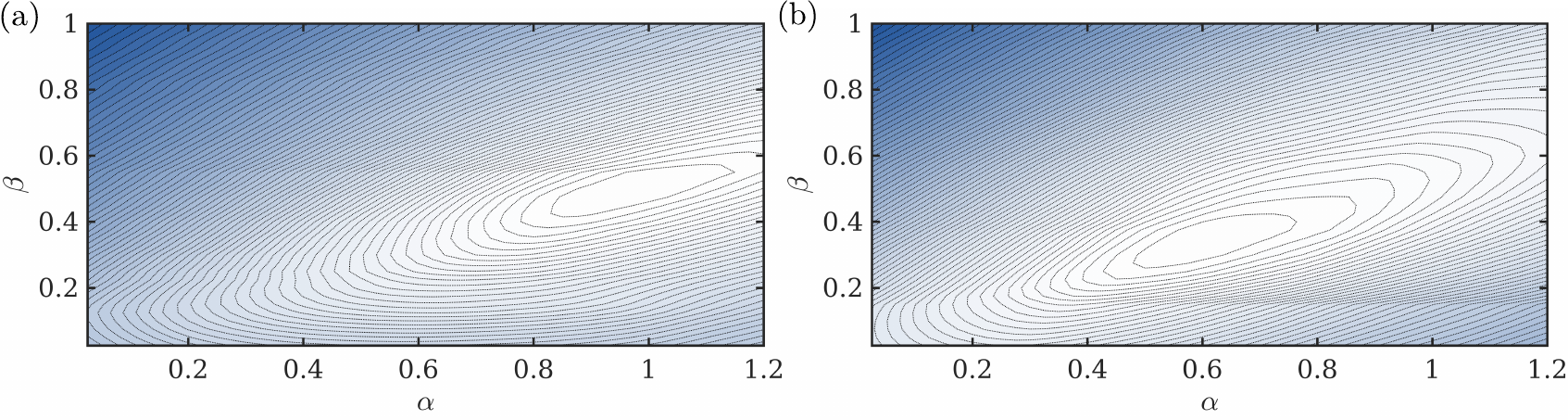}
\vspace{-2mm}
\caption{Contour plot of the error $E$ of the self-similar fitting, see Eq.\,(\ref{eq:error}), as a function of the parameters $(\alpha,\beta)$ for (a) short times in the Gaussian regime and (b) long times in the non-gaussian regime. 
Lighter colors indicate decreasing errors. In panel (a), we used datapoints of the distribution function in a time window of size $\Delta t = 10\tau_*$ centered at $t_0 = 15\tau_*$. In panel (b), we used datapoints for the distribution function in a time window of size $\Delta t = 10\tau_*$ centered at $t_0 = 25\tau_*$. }
\label{fig:errors}
\end{figure}

\section{IV. The XXZ model}

The central ingredients in the discussion of the main text are the global SU(2) symmetry of the Heisenberg Hamiltonian and the dimensionality $d=2$. We checked that the XXZ model exhibits different scaling behavior as we tune the anisotropic exchange $J_z$ from easy-plane to easy-axis across the isotropic point. The XXZ model is given by
\be
\hat{H} = -\sum_{\langle i,j \rangle} \left[ J(S_{i}^+S_j^-+S_i^-S_j^+) + J_z S_i^zS_j^z \right], 
\ee
which, in the low spin wave density limit, can be interpreted as a Bose gas in which the $J(\hat{S}_{i}^{+}\hat{S}_j^{-}+{h.c.})$ terms give rise to hopping and the $J_zS_{i}^zS_{j}^z$ terms give rise to interactions. In the regime $J_z\ll J$, we expect to see the same scaling behavior as that observed in a weakly-coupled Bose gas ($\beta = 1/2$, $\alpha = d\beta$). Our numerical simulations for $J_z/J_\perp = 0.5$ show exactly these scaling exponents, see Fig.~\ref{fig:betat}(a-b). 
  
\section{V. Equations of motion}

In the main text, we used the equations of motion of spin operators to analytically derive the dynamical scaling exponents. This approach is far more general than kinetic theory which may not be legitimate if, for example, magnetization fluctuations are large (as in our case). The microscopic equations of motion (in units of $J$) are given by:
\be
\partial_t \hat{S}_{i}^a = \epsilon_{abc}\sum_j \hat{S}_i^b \hat{S}_j^c, 
\ee
where we sum over repeated indices only if the appear both as subscripts and superscripts. Going to momentum space, we obtain
\be
\partial_t \hat{S}_{\bm k}^a = \left( \gamma_0 - \gamma_{\bm p}\right)\epsilon_{abc} \hat{S}_{\bm k - \bm p}^b \hat{S}_{\bm p}^c + i \gamma_0 S_{\bm k}^a,  
\ee
with $\gamma_{\bm k} = \sum_{\bm \ell}e^{i{\bm k}\cdot{\bm \ell}}$ (${\bm \ell}$ are unit cell vectors). Multiplying on the left with the operator $\hat{S}^a_{-\bm k}$, summing  with the complex conjugate of $\hat{S}_{-\bm k}^a\partial_t\hat{S}_{\bm k}^a$, and taking expectation value, results in 
\be
\partial_t \langle \hat{S}_{-\bm k}^a \hat{S}_{\bm k}^a \rangle = 2 \sum_{\bm p} (\gamma_0-\gamma_{\bm p}){\rm Re}\left[\epsilon_{abc}\langle \hat{S}_{-\bm k}^a\hat{S}_{\bm k - \bm p}^b\hat{S}_{\bm p}^c\rangle\right].
\label{eq:S2eom_SM}
\ee
We emphasize that correlations of the form $\langle \hat{S}_{\bm k}^x\hat{S}_{\bm p}^y\hat{S}_{\bm q}^z\rangle$ appearing on the right-hand side of Eq.~(\ref{eq:S2eom_SM}) are not necessarily zero because the three components of magnetization are not independent. Equation (\ref{eq:S2eom_SM}) is used in the main text to derive the scaling exponents under the assumption that a single lengthscale $\xi$ (the quasi-condensate correlation length) describes collective dynamics and correlation functions at intermediate timescales. 

\section{VI. Effective Hamiltonian and Gaussian fixed point}

At short times, we observe a scaling regime with exponents $(\alpha,\beta)\approx (1,1/2)$, which we attribute to a Gaussian fixed point. 
In addition, we argue in the main text that non-linearities in the Heisenberg model are marginal in $d=2$. Both observations can be rationalized from a Holstein-Primakoff expansion of the Heisenberg model close to the ferromagnetic ground state. Assuming small deviations from the ferromagnetic ground state $|F\rangle$ (all spins pointing up) and using the Holstein-Primakoff transformation, $\hat{S}_j^+ = \sqrt{2S-\hat{\psi}_{j}^\dagger\hat{\psi}_{j}}\hat{\psi}_j$ and $\hat{S}_j^z = S -\hat{\psi}_j^\dagger\hat{\psi}_j$, to quartic order in the bosonic operators $\hat{\psi}_j$ leads to the long-wavelength Hamiltonian
\be
\hat{H} = JSa^2 \int_{\bm x} \left(\nabla\hat{\psi}_{\bm x}^\dagger \nabla\hat{\psi}_{\bm x} + \frac{1}{4S}\hat{\psi}_{\bm x}^\dagger\hat{\psi}_{\bm x}^\dagger\nabla\hat{\psi}_{\bm x}\nabla\hat{\psi}_{\bm x}+{\it h.c.}\right).
\label{eq:Heffective}
\ee
Unlike the usual Bose gas with hard core collisions, here the collision amplitude of two quasiparticles with momentum ${\bm k}$ and ${\bm p}$ is $\propto - ({\bm k}\cdot{\bm p})$. This reflects the SU(2) symmetry of the Hamiltonian: collisions become negligible at small momenta because a ${\bm k}\rightarrow 0$ magnon state, $\hat{\psi}_{\bm k}^\dagger|{\rm F}\rangle \approx \frac{\hat{S}_{\bm k}^-}{\sqrt{2S}}|{\rm F}\rangle$, is effectively a global rotation of $|{\rm F}\rangle$ that would not affect the dynamics of a second incoming magnon. 

The Gaussian exponents discussed in the main text can be derived by dropping non-linearities from~\eqref{eq:Heffective}, and scaling the Fourier transform of the field with $\xi$ as in Eq.~(7), following $\psi_\mathbf{k}\sim \xi^{\alpha/(2\beta)}$ and $\xi\sim t^{\beta}$ of the main text.
We can now take the  action associated to the free version of~\eqref{eq:Heffective} (equivalent to considering just the kinetic term), and require that this is invariant under a running scale, $\xi$. It will yield     $\xi^{1/\beta-d-2+\alpha/\beta}\sim \xi^0$ or, in other words, $(\alpha+1)/\beta=d+2$.
Combining it with   relation (6)  in the main text, $\alpha=\beta d$, related to total spin conservation,  we find $\alpha=1$ and $\beta=1/2$ in $d=2$.  Such exponents are observed at short times in Fig.~\ref{fig:betat}. Note that one can also derive the same Gaussian exponents in a similar way as for the non-thermal fixed point in Eq.~(7) by deriving the equations of motion associated to Eq.~(\ref{eq:Heffective}) without non-linearities, $\partial_t \hat{\psi}_{\bm k} \sim |{\bm k}|^2 \hat{\psi}_{\bm k}$, and rescaling both sides, which directly yields $\beta=1/2$.

\begin{figure}
\centering\includegraphics[scale = 1.0]{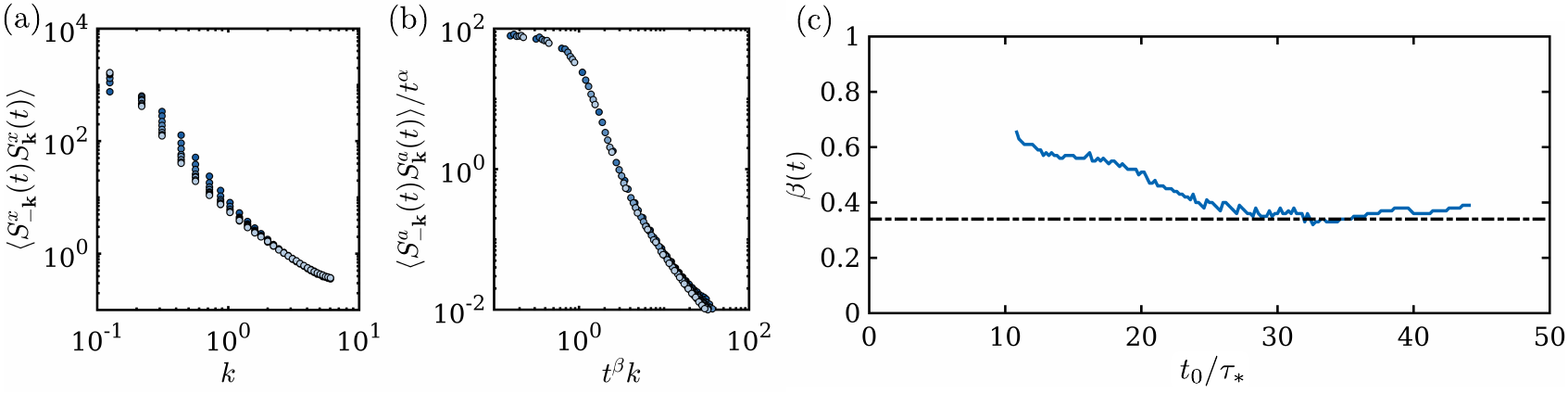}
\vspace{-2mm}
\caption{(a-b) Equal-time spin-spin correlation function for the the easy-plane XXZ model. Panel (a) shows the unrescaled data whereas panel (b) shows the rescaled data. Simulation parameters: $q_x a = 0.5)$, $q_y = 0$, $\theta = \pi/4$, $J_z/J = 0.5$. 
(c) Crossover between the Gaussian fixed point ($\beta\approx 0.5$; plateau in the time window $10\lesssim t_0/\tau_* \lesssim18$) and the non-thermal fixed point ($\beta\approx 1/3$; plateau starting at $t_0/\tau_*\gtrsim 25$). Here the exponents $(\alpha,\beta)$ are computed in a time window of size $\Delta t = 10\tau_*$ centered at different values of $t_0$. We do not plot data below $t_0<10\tau_*$ as 
a macroscopic occupation of the spin mode with momentum $\bm q$ still remains.}
\label{fig:betat}
\end{figure}

\section{VII. Soft vertices in the Holstein-Primakoff field theory \\ of the Heisenberg model}

The leading non-linearity in the low-density Holstein-Primakoff  expansion of the Heisenberg magnet is a `soft' vertex~\cite{rodriguez2020hydrodynamic,halperin1969hydrodynamic}, a scattering term ruled by spatial gradients which would therefore vanish at low momenta in Fourier space.
In Keldysh non-equilibrium field theory~\cite{kamenev2011field}, this    appears as a quartic term $\propto \lambda\int_{t,x}  \nabla^2\varphi^3\tilde{\varphi}$, while for $O(n)$ models we have the `hard' vertex $\propto g\int_{t,x}\varphi^3\tilde{\varphi}$. Here, the fields $\varphi$ and $\tilde{\varphi}$ incorporate classical and quantum fluctuations, see for instance~\cite{polkovnikov2010phase,kamenev2011field,2015bergesreview}. The Laplacian $\nabla^2$ is responsible for the soft vertex proportional to momentum squared in Eq.~\eqref{eq:S2eom}, and it has been  originally derived within   spin-wave hydrodynamics~\cite{rodriguez2020hydrodynamic,halperin1969hydrodynamic}. At the level of canonical power counting, the two vertices  scale respectively as $\lambda\sim\xi^{-(2-d)}$ and $g\sim\xi^{-(4-d)}$ in a regime dominated by classical  fluctuations; this regime  is justified by the high energy of the initial states we use in this work. As a result, a different power counting is obtained for higher order non-linearities: they become less relevant with higher powers of $\varphi$ for $O(n)$ models in $d=2$, while they are all marginal in the Heisenberg magnet. Therefore, the effective field theory for $O(n)$ models and the Heisenberg magnet differ already  at the level of canonical power counting, and this in turn dictates  a different set of exponents for dynamical  scaling, whenever diagrammatic corrections are included. This hints that the two models belong to different non-equilibrium universality classes. 

\end{widetext}

\end{document}